\documentclass[10pt, letterpaper]{article}

\usepackage{cogsci}
\usepackage{pslatex}
\usepackage{float}
\usepackage{caption}

\usepackage{amsmath}

\usepackage{amssymb}

\usepackage{hyperref}

\usepackage{graphicx}

\usepackage{fancyvrb}
\DefineVerbatimEnvironment{Sinput}{Verbatim}{fontshape=sl}
\DefineVerbatimEnvironment{Soutput}{Verbatim}{}
\DefineVerbatimEnvironment{Scode}{Verbatim}{fontshape=sl}

\DefineVerbatimEnvironment{Code}{Verbatim}{}
\DefineVerbatimEnvironment{CodeInput}{Verbatim}{fontshape=sl}
\DefineVerbatimEnvironment{CodeOutput}{Verbatim}{}
\newenvironment{CodeChunk}{}{}

\usepackage{cite}

\usepackage{color}

\title{Preserved Structure Across Vector Space Representations}

\usepackage[utf8]{inputenc}
\usepackage[export]{adjustbox}

\author{{\large \bf Andrei Amatuni, Estelle He, Elika Bergelson} \\ \texttt{\{andrei.amatuni, estelle.he, elika.bergelson\}@duke.edu} \\ 417 Chapel Dr. Durham, NC 27708 USA \\ Department of Psychology and Neuroscience \\ Duke University}

\begin{document}

\maketitle

\begin{abstract}
Certain concepts, words, and images are intuitively more similar than
others (dog vs.~cat, dog vs.~spoon), though quantifying such similarity
is notoriously difficult. Indeed, this kind of computation is likely a
critical part of learning the category boundaries for words within a
given language. Here, we use a set of 27 items (e.g. `dog') that are
highly common in infants' input, and use both image- and word-based
algorithms to independently compute similarity among them. We find three
key results. First, the pairwise item similarities derived within
image-space and word-space are correlated, suggesting preserved
structure among these extremely different representational formats.
Second, the closest `neighbors' for each item, within each space, showed
significant overlap (e.g.~both found `egg' as a neighbor of `apple').
Third, items with the most overlapping neighbors are later-learned by
infants and toddlers. We conclude that this approach, which does not
rely on human ratings of similarity, may nevertheless reflect stable
within-class structure across these two spaces. We speculate that such
invariance might aid lexical acquisition, by serving as an informative
marker of category boundaries.

\textbf{Keywords:}
vector space models; semantic similarity; word learning
\end{abstract}

\hypertarget{introduction}{%
\section{Introduction}\label{introduction}}

Infants are presented with a challenge to carve the world into distinct
lexical entities in the process of learning their first language.
They're provided with little supervision while mapping a territory that
William James (1890) famously dubbed a ``great blooming, buzzing
confusion''. How they determine which aspects of the world to attend to
in service of this goal is an area of ongoing research and debate
(Mareschal \& Quinn, 2001). Relatedly, features of objects and their
environments are varyingly informative with regards to object
segmentation and category structure. Some researchers have suggested
that categorization is along fundamentally perceptual grounds and that
only later in development is conceptual knowledge incorporated into
these nascent perceptual categories (Quinn \& Eimas, 1997, 2000; Quinn
et al., 2000). Others suggest that there are in fact two distinct
processes at work, such that perceptual categories are computed
automatically by the sensory systems, while conceptual categories are
independently formed through conscious action (Mandler, 2000). Träuble
and Pauen (2007) provide evidence of functional information (regarding
the animacy of objects) influencing early category judgements. Gelman
and Markman (1986) explicitly set these two sources of category cues
against each other (i.e.~functional vs.~perceptual), and find that
preschoolers can override perceptual overlap in reasoning about
functional similarity in natural kinds.

The degree to which conceptual and perceptual information are separable
in early learning and in adult experts is an important open question.
Any model which hopes to explain the mechanics of human categorization
must address how potentially disparate information-sources interface in
mental representations, and to what degree they interact. Indeed,
evidence from human learners suggests they integrate perceptual and
linguistic information during categorization and learning (Colunga \&
Smith, 2005; Sloutsky, 2003; Sloutsky, Lo, \& Fisher, 2001). Here we
take on a deliberately different approach. We separate computations over
images and words, and then compare the overlap in the similarity among
items that these systems deduce. Using a set of highly familiar and
common words and concepts from a large infant corpus, we compare the
output of an image-based similarity analysis and a word co-occurrence
similarity analysis for these same items. We use algorithms that learn
feature representations without hand engineering, purely as a byproduct
of their separate training objectives (i.e.~natural language processing
vs.~object recognition). Comparing the representations these algorithms
learn provides a window into the structure of visual and semantic forms.

The terminology in this area of research can be challenging. Delineating
the differences between words, concepts, and categories in the abstract,
and the processes which underlie identifying, understanding, or
comparing particular instances of them is not trivial. For present
purposes, we stick to concrete, `basic level' nouns that are
early-acquired, since our underlying question concerns how such words
are learned. We assume that nouns refer to concepts, which have
categorical boundaries (such that cats are not in the `dog' category),
while acknowledging that multiple nouns can refer to a given concept,
and different concepts can be called to mind by a given word. We further
assume that specific instances of words and specific referents of the
concept a word picks out are both used to learn the word's meaning and
the concept's category boundaries. We use the term `item' to refer to
the words/concepts we examine.

Intuitively, word similarity and image similarity are likely to overlap
to some degree, since they describe the same underlying entity. Here we
explore whether the similarity spaces generated by two disparate
algorithms give rise to \emph{similar} similarities among high-frequency
items. If they do, it supports the notion of an underlying invariance
across representational formats that is capturable by these models. We
further examine whether the same ``neighboring'' items are picked out
within these two spaces. One might imagine that the properties that
render images similar and words similar are different enough that the
overlap will be minimal; in contrast, high overlap would again suggest a
true invariance being captured by both word- and image-tokens. Finally,
we examine whether having more neighbors within word- and image-space
influences early learning. Given that similarity makes word-learning and
category-learning more difficult (Rosch \& Lloyd, 1978; Stager \&
Werker, 1997), we hypothesize that items with more neighbors will be
later-learned (i.e.~known by fewer children of a given age.)

\hypertarget{methods}{%
\section{Methods}\label{methods}}

\hypertarget{items}{%
\subsection{Items}\label{items}}

We analyze 27 high-frequency items from infants' early visual and
linguistic input, aggregated as part of SEEDLingS, a project including
longitudinal audio and video data of infants' home environments from
6-17 months (Bergelson, 2016a, 2016b). We briefly describe this larger
study to relay how these 27 items were chosen. In the larger study, 44
infants were tested every other month (from 6-18 months) on common
nouns, using a looking-while-listening eyetracking design in which two
images are shown on a screen and one is named. The words for these
experiments were chosen by dint of being high frequency or well known
across infants in other samples, e.g.~the Brent Corpus and WordBank
(Brent \& Siskind, 2001; Frank, Braginsky, Yurovsky, \& Marchman, 2017),
or by being in the top 10 concrete nouns heard in each infant's own home
recordings in the preceding two months.

The images displayed when these words were tested were chosen from a
library of prototypical images (e.g.~dog) and images of infants' own
items, as seen in their home videos (e.g.~a given infant's cat, specific
bottle, etc.). To enter the current analysis, images had to occur
\textgreater{}9 times in this image library of high frequency concrete
nouns derived from 264 eyetracking sessions (image counts:
M=18.85(10.89)). These words were heard extremely often over the 528
daylong audio-recordings and 528 hour-long video recordings of these 44
infants (M=2034.85(1540.22)). Thus, the words and images used here
provide an ecologically-valid item-set for present modeling purposes.

The images of the 27 items used to derive average category image-vectors
were all 960x960 pixel photos of a single object on a gray background.
Items correspond to words found on WordBank (Frank et al., 2017), a
compilation MacArthur-Bates Communicative Development Inventories, used
as a proxy for age of acquisition below (Dale \& Fenson, 1996).

\hypertarget{vector-representations}{%
\subsection{Vector Representations}\label{vector-representations}}

We generate two sets of vector representations for these early-learned
items. The first set is taken from a pretrained GloVe representations
(Pennington, Socher, \& Manning, 2014), a modern distributional semantic
vector space model. The second is taken from the final layer activations
of a pretrained image recognition model, Google's Inception V3 CNN
(Szegedy, Vanhoucke, Ioffe, Shlens, \& Wojna, 2016). Both of these
representations are generally referred to as ``embeddings''. They map
objects from one medium (e.g.~images or words) into a metric space where
distances between points can be computed and function as similarity
measures. All code used for generating these vectors and the subsequent
analysis can be found on
Github.\footnote{\url{https://github.com/BergelsonLab/preserved_structure}}

\hypertarget{word-vectors}{%
\subsubsection{Word Vectors}\label{word-vectors}}

Our word vectors are based on GloVe's instantiation of the
distributional hypothesis: co-occurring words share similar meaning
(Firth, 1957; Harris, 1954). Thus, by capturing the covariance of tokens
in large corpora, we can capture aspects of semantic structure. We use
the set of vectors pretrained by the GloVe authors on the Common Crawl
corpus with 42 billion tokens, resulting in 300 dimensional vectors for
1.9 million unique
words\footnote{\url{https://nlp.stanford.edu/projects/glove/}}. Such
vectors have shown promise in modeling early semantic networks (Amatuni
\& Bergelson, 2017). Thus, in word vector space (hereafter word-space),
each of our 27 items is represented as a 300-dimensional vector, with
each word assigned a unique point in a common vector
space.\footnote{While this corpus is best-suited to our goal of modelling 'how words behave' writ large, we also conducted the analyses below with vectors trained on the North American English CHILDES corpora (MacWhinney, 2000), which is $\sim$4000x smaller. We observe the same qualitative patterns.}

\hypertarget{image-vectors}{%
\subsubsection{Image Vectors}\label{image-vectors}}

The image embeddings are taken from the final layer of activations in a
CNN, whose objective function tunes network parameters in service of
object recognition, computing loss in reference to labeled training
images. These tuned parameters determine the value of our vectors,
transforming the input image signal as it passes through the network.
The final layer of this network encodes the most abstract and integrated
visual features, serving as the basis for classification into 1000
different classes. The model was trained on the ILSVRC-2012-CLS
challenge dataset, which defined the 1000 ImageNet category subset
(Russakovsky et al., 2015).

Unlike the word vectors we use, different images containing the same
type of item will have varying vector representations after passing
through the layers of a neural network. This presents a problem in
comparing the two forms of representation. Thus, we first define the
most prototypical image vector for any given category of object, which
will generate our 2048-dimensional representation for each of the 27
items, in image vector space (hereafter image-space).

Given a set of images \(S_c\) containing objects belonging to a single
category \(c\) (e.g.~cat, chair), we define our prototypical vector
\(\hat{x}_c\) of \(S_c\) as the generalized median within a
representational space \(U\). This is the vector with minimal sum of
distances between it and all the other members of set \(S_c\) in \(U\).
If \(x\) and \(y\) are vectors in space \(U\), products of images in
\(S_c\) being passed through a neural network, then
\(\hat{x}_c = \operatorname*{arg\,min}_{x\in U} \sum_{y\in U} d(x, y)\).
We define \(d(x, y)\) as the cosine distance measure:
\(d(x, y) = 1 - \frac{x\cdot y}{\|x\|\|y\|}\)

This is not a distance function in the strict sense, but unlike
Euclidean distance, is less susceptible to differences in \(L^2\) norm
influencing our measure of similarity. Thus in principle, cosine
similarity corrects for frequency effects inherent to the training data.
To validate our image vectors, we benchmarked classification accuracy,
finding that Inception V3 is indeed learning useful representations of
these highly child-centric images: the model's top prediction was of the
correct item-class 26\% of the time; in 63\% of cases it was in the top
5. Even when incorrect, predictions tended to reflect idiosyncrasies in
child-relevant items, e.g.~the top guess for a cartoon puppy was
``teddy''.

\hypertarget{comparing-spaces}{%
\subsection{Comparing spaces}\label{comparing-spaces}}

Having computed our two sets of vectors (i.e.~word-space and
image-space), we can compare pairwise distances between items, both
within a single space and across the two. When comparing across the two
spaces, a correlation in pairwise distances implies that inter-object
distances have been conserved. For example, if ``dog'' and ``cat'' are
close together in word space and mutually far apart from ``chair'' and
``table'' in that same space, maintaining this relationship for all
pairwise distances in the \textit{other} vector space means that the
global inter-object structure is preserved across this mapping. This is
despite being in strikingly different spaces, both in terms of
dimensionality (words:300, images:2048) and by virtue of using
completely different algorithms and inputs to establish the vector
representations for items. So while their absolute locations might have
been transformed, this correlation (and related computations) would be a
measure of the \textit{degree of invariance} in their positioning
relative to each other.

\begin{CodeChunk}
\begin{figure}[tb]
\includegraphics{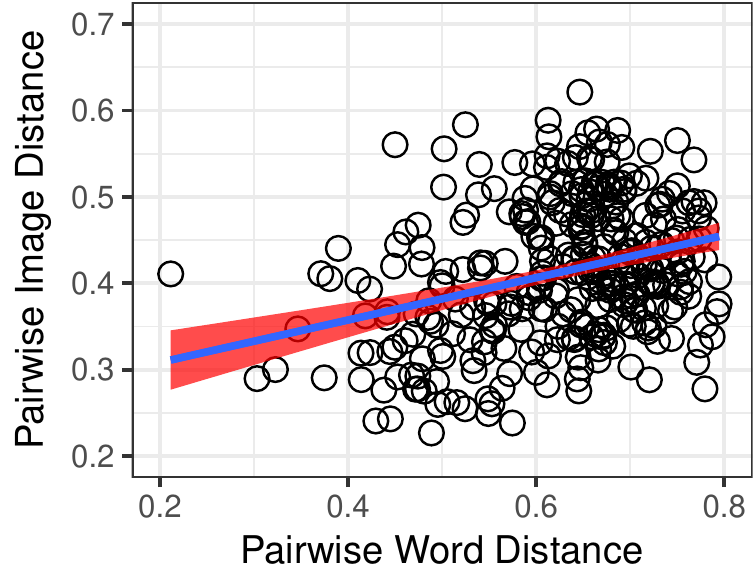} \caption[Relative cosine distance between items in word-space (x-axis) and image-space (y-axis), for each item pair]{Relative cosine distance between items in word-space (x-axis) and image-space (y-axis), for each item pair. Fitted line reflects linear fit with SE ($R = 0.30$, $p < .001$).}\label{fig:pairwise-corr}
\end{figure}
\end{CodeChunk}

\hypertarget{results}{%
\section{Results}\label{results}}

To test whether image- and word-based similarity converged, we conducted
several analyses. First, we tested whether the pairwise cosine distances
for all items in word-space correlated with those same pairwise
distances in image-space (see Figure \ref{fig:pairwise-corr}). We find a
significant correlation among the 351 pairs of distances (\(R = 0.30\),
\(p < 1.5e-08\))\footnote{since distances are identical for cat-dog and dog-cat, and since we omit an item's distance to itself (0), there are (27*27-27)/2) pairs of distances. For simplicity, we report Pearson's R and plot linear fit on Fig. 1; non-parametric correlations (e.g. Spearman's $\rho$) reveal the same pattern.}.

Next, we examined the degree to which our set of 27 words shared
overlapping `neighbors' in the two vector spaces (see Table
\ref{tbl:overlap-table}). We defined neighbor by first determining the
mean similarity distance between each item and the 26 other items. Any
items whose distance to this target had a z-score of less than -1 was
considered a neighbor. Within word-space, items had on average 4.26
neighbors (\(SD=1.38\) , \(R=2-7)\). Within image-space, items had 2.51
neighbors (\(SD=1.6\), \(R=1-6\)).

We next tested whether both spaces picked out overlapping neighbors
(e.g.~whether the neighbor of `cat' in image-space overlapped with the
neighbors of `cat' in word-space. The majority of items have at least 1
neighbor which is shared across representational spaces. We quantified
this through overlap ratios: (\# overlap)/(\# neighbors). Overlap was
significantly greater than 0 (\(M=0.165\), \(SD=0.16\), \(p < 0.001\) by
Wilcoxon test). This complements the correlational analysis, showing not
only do the distances for any given pair tend to have similar values in
image- and word-space, but that the \emph{most} similar words/images
(i.e.~each item's neighbors) were also consistent across these spaces.

\hypertarget{connecting-with-learnability}{%
\subsection{Connecting with
Learnability}\label{connecting-with-learnability}}

While some degree of convergence across image and word spaces is
expected given that these are two different manifestations of the same
underlying concept/word/item, we next queried whether this invariance
related to learnability. We hypothesized that words with more
overlapping neighbors would be harder for children to learn, since both
the visual and linguistic spaces they occur in are more `cluttered.' To
test this, we looked at the relative rates of acquisition of these items
in WordBank (Frank et al., 2017), using the 6945 children's data from
English. Since we did not have clear predictions about specific ages, or
of tradeoffs between comprehension and production, we used both. I.e.,
we used comprehension norms (from MCDI-Words and Gestures, averaging
over 8-18 months) and production norms (from MCDI-Words and Sentences,
averaging over 16-30 months).

\begin{CodeChunk}
\begin{figure}[tb]
\includegraphics{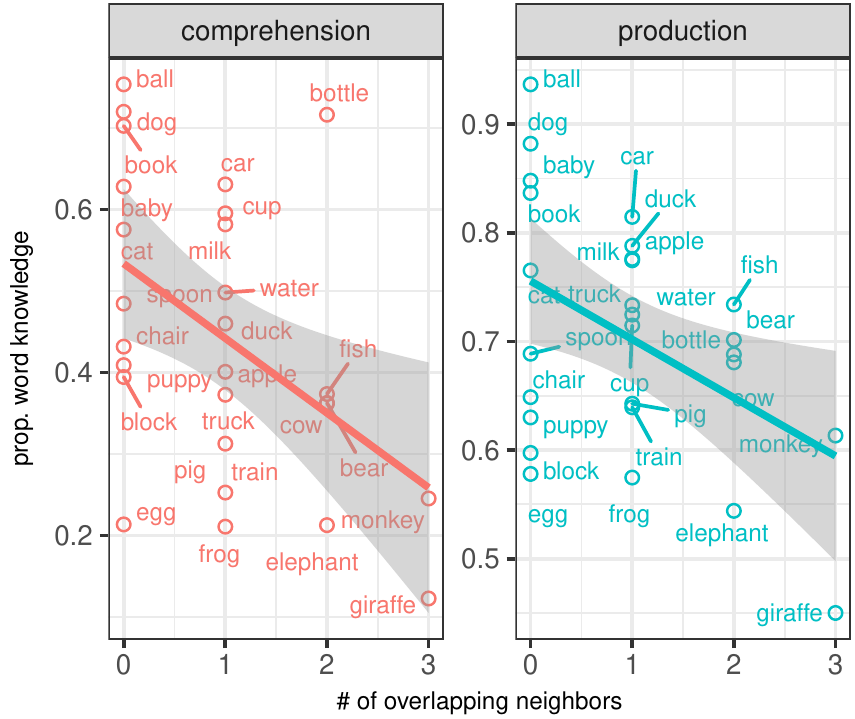} \caption[Proportion of children in WordBank reported to understand (left, 8-18mo.average) or produce (right, 16-30mo]{Proportion of children in WordBank reported to understand (left, 8-18mo.average) or produce (right, 16-30mo. average) the 27 items, as a function of number of overlapping (image and word) neighbors. Lines indicates linear fit with SE CIs (both R$>$-.45, p$<$.05).}\label{fig:overlap-aoa-graphs}
\end{figure}
\end{CodeChunk}

We found that the number of overlapping neighbors a given word had was
negatively correlated with the proportion of children who were reported
to understand (\(R=-0.48\), \(p = 0.011\)) and produce the word
(\(R=-0.46\), \(p = 0.017\)); see Figure \ref{fig:overlap-aoa-graphs}.
That is, words with more overlapping neighbors are later-learned
(i.e.~known by fewer children) than words with fewer overlapping
neighbors. To test whether this was specific to overlap, we examined the
number of image-only and word-only neighbors (see Table
\ref{tbl:overlap-table})), but found no correlations with word knowledge
(all \(p > .05\)).

Analyses using randomly generated vectors of identical dimensionality
showed no preserved structure across spaces, a significantly smaller
overlap ratio (\(M\) = 0.048, \(SD=0.074\), \(p = 0.006\) by Wilcoxon
test), and critically, no correlations with learning.

\begin{CodeChunk}
\begin{table}[tb]
\includegraphics[width=1\linewidth]{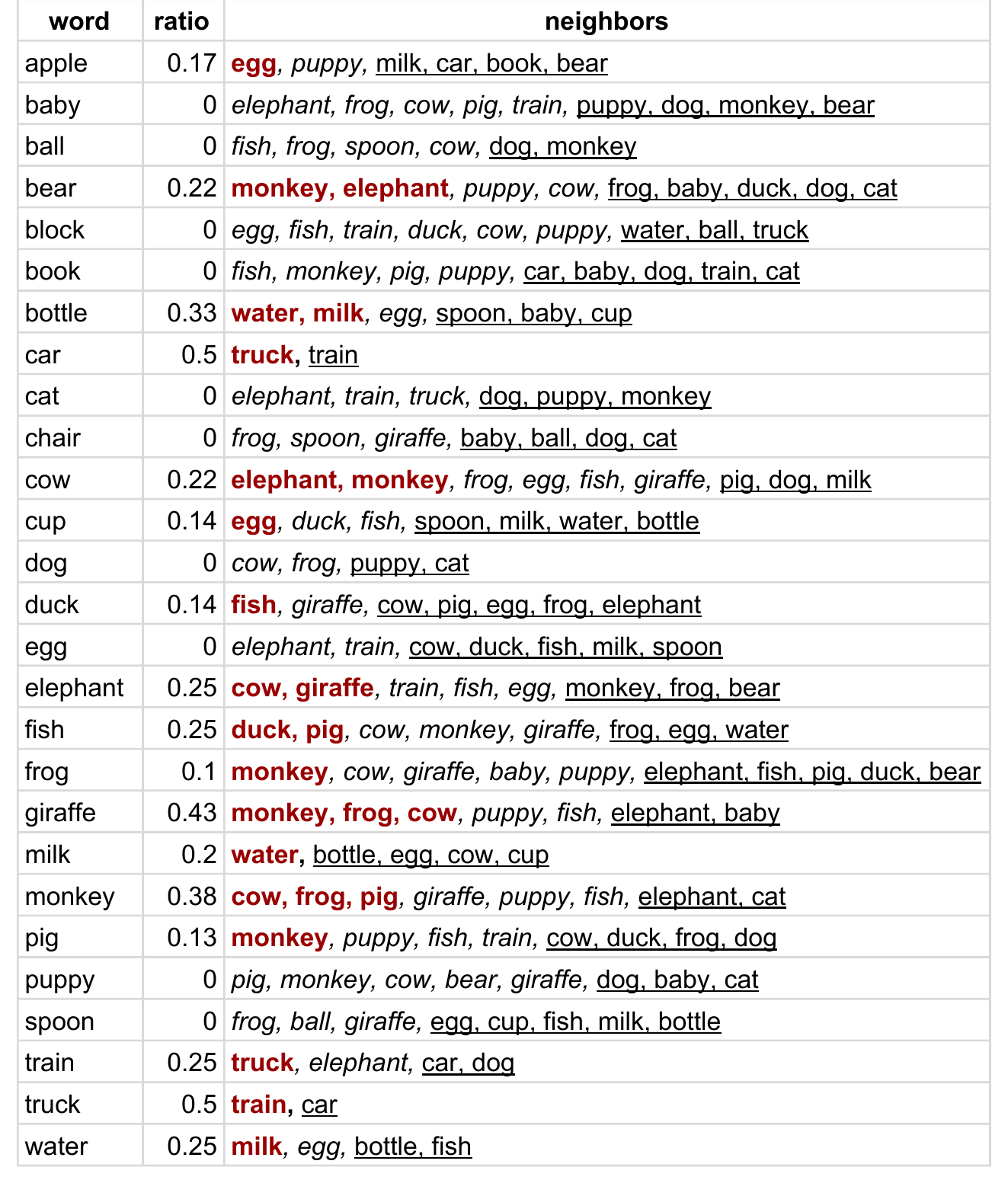} \caption[Neighbors in image- and word-space]{Neighbors in image- and word-space. Overlapping neighbors are in bold and red; italicised words are image-neighbors only, underlined words are word-neighbors only. Overlap ratio reflects shared over total neighbors.}\label{tbl:overlap-table}
\end{table}
\end{CodeChunk}

\hypertarget{discussion}{%
\section{Discussion}\label{discussion}}

The results above revealed a notable correspondence between
representations learned by two different algorithms operating over
inputs in two fundamentally different encodings (i.e.~visual and
linguistic). We find that not only are the relative distances among
these 27 common, early-learned items correlated across word- and
image-space, but that even at the item-level, the closest words
(i.e.~neighbors) in both spaces overlap as well. Moreover, words
corresponding to items with more neighbors were reportedly less well
known by young children. Notably, the common ground between these
representations is the real life concepts they both aim to model. This
is particularly noteworthy given the dimensionality of our feature
spaces, and that these algorithms were placed under no pressure to find
homologous representations. That said, we do not suggest that these
algorithms learn representations the way children do, though we note
that CNN architectures were originally inspired by primary visual cortex
(Fukushima \& Miyake, 1982) and GloVe has recently been used to decode
semantic representations from brain activity (Pereira et al., 2018). To
be clear, none of the corpora used to train the models here are meant to
represent a \emph{given} child's input or representations. They are
meant to study structure inherent \emph{in the learning data}, namely
the degree to which visual and linguistic information is purely
separable.

The notion that we can make inferences about one aspect of an object
given another aspect is not surprising or controversial. Rather than
considering multiple dimensions at once as real learners must, here we
show that even with the experiences parcelled out separately into visual
and linguistic spaces, `similarity' is conserved to some degree. That
said, a limitation of the current work is that the image vectors are
trained on images in context; while in real life images occur with
informative context, extending this approach to decontextualized images
using unsupervised algorithms would provide a cleaner demonstration of
`purely' visual similarity; we save this for future work. Relatedly,
while we used common infant-oriented items, a wider item-space (less
skewed to animals for instance) would fruitfully extend this research.

Through what metrics can a learning algorithm, or indeed a human,
establish gradations of likeness? Are these necessarily the same metrics
which form the basis of category boundaries? These are fundamental
questions in the field (Edelman, 1998; Hahn, Chater, \& Richardson,
2003; Kemp, Bernstein, \& Tenenbaum, 2005; Shepard \& Chipman, 1970;
Tversky, 1977). While our current results are not sufficient to support
a specific mechanism, it suggests a special role for invariance, given
that the unifying thread between our algorithms and inputs are the
common objects they represent. Underneath the diversity of visual
statistics and token distributions lie stable entities in the world
which give rise to regularity across measurements at different vantage
points (i.e.~modalities), an idea dating back to Helmholtz (1878).
Recent findings examining the mechanics of generalization in DNNs lend
modern information theoretic support to this notion (Achille \& Soatto,
2017; Shwartz-Ziv \& Tishby, 2017). That said, many things that `go
together' are not visually similar, but rather have hierarchical,
functional, or associative relations (e.g.~carrot/vegetable,
skateboard/boat, carrot/bunny, respectively); we leave this to future
work.

In principle, one would expect that words with greater invariance across
different representational dimensions would be learned earlier, since
such representations are likely easier to make categorical inferences
over. Indeed, while many words lack visual correlates, e.g.~grammatical
markers or unobservables (Gleitman, Cassidy, Nappa, Papafragou, \&
Trueswell, 2005), words with less consistent visual features are
generally learned later (Bergelson \& Swingley, 2013; Dale \& Fenson,
1996). This is in contrast to the more helpful scenario where, for
example, visibly round objects occur with `roll'; indeed such correlated
perceptual and linguistic cues aid the child learner (Yoshida \& Smith,
2005).

Our WordBank-based analysis speaks to this, highlighting that for these
concrete nouns, when the space is cluttered along both visual and
linguistic dimensions, learning is slower. Notably, we find no effect of
word- or image-only neighbors, suggesting that the overlap-neighbor
effect may indeed be due to the (perhaps more noticeable) ``clutter''
across spaces. Such an account is in keeping with research that finds
that for displays of semantically-similar items, word comprehension is
reduced (Arias-Trejo \& Plunkett, 2010; Bergelson \& Aslin, 2017).
Similar feature-space cluttering effects have been a standard in the
visual search literature, where target-to-distractor similarity reduces
speed and accuracy of search (Duncan \& Humphreys, 1989; Treisman \&
Gormican, 1988). Interestingly, when learning the structure of new
objects, adults learning Chinese characters perform better in visual
search when they are trained with characters that force attention to the
points of confluence between features (Popov \& Reder, 2017); perhaps
infants do the same. That said, while concrete nouns are an appropriate
target for these current analyses given their demonstrably early age of
acquisition (Dale \& Fenson, 1996), further work with more abstract
words (nouns and beyond) is a clear next step.

The results here provide in-principle proof that vector space models of
words and images can be fruitfully combined and linked to early language
and concept learning. Our approach could readily be extended to examine
learning-relevant properties like animacy, shape, and color (e.g.~Frank,
Vul, \& Johnson (2009), Landau, Smith, \& Jones (1992)). Along these
lines, in an exploratory analyses with these items we find that
splitting item-pairs into animate, inanimate, and mixed categories
suggests that the image- and word-based correlation is particularly
strong for the inanimate items, though with only 27 items, further
conclusions are as-yet unwarranted. However, one can imagine that in a
larger database of children's visual and linguistic experiences, tests
of overlapping similarity and relative degrees of within-class structure
conservation may provide informative leverage for predicting age of
acquisition across the early vocabulary.

\hypertarget{conclusion}{%
\section{Conclusion}\label{conclusion}}

We find evidence of links between visual and linguistic features learned
by two distinct machine learning algorithms which operate over
drastically different inputs, and are trained in the service of
unrelated ends. These links suggest conserved structure between these
two separable information sources (i.e.~images and words). Indeed, it
seems that not only do these algorithms converge on which items are
`closer' in similarity within a group of oft-heard and seen concrete
nouns, but that children are sensitive to these overlapping cross-word
relationships as well. The process that created the word- and
image-spaces we examine here is certainly not meant to be cognitively
plausible. Nevertheless, our results suggest that this vector-space
approach can be tied to language acquisition, and provides promising new
avenues for uncovering cross-representational influences on early word
and concept learning.

\hypertarget{acknowledgements}{%
\section{Acknowledgements}\label{acknowledgements}}

We thank the BLAB, Eric Bigelow, \& NIH DP5-OD019812.

\hypertarget{references}{%
\section{References}\label{references}}

\setlength{\parindent}{-0.1in} 
\setlength{\leftskip}{0.125in}

\noindent

\hypertarget{refs}{}
\leavevmode\hypertarget{ref-achille2017emergence}{}%
Achille, A., \& Soatto, S. (2017). On the emergence of invariance and
disentangling in deep representations. \emph{arXiv Preprint
arXiv:1706.01350}.

\leavevmode\hypertarget{ref-amatuni2017semantic}{}%
Amatuni, A., \& Bergelson, E. (2017). Semantic networks generated from
early linguistic input. In \emph{Proceedings of the 39th annual
conference of the cognitive science society} (pp. 1538--1543).

\leavevmode\hypertarget{ref-arias2010effects}{}%
Arias-Trejo, N., \& Plunkett, K. (2010). The effects of perceptual
similarity and category membership on early word-referent
identification. \emph{Journal of Experimental Child Psychology},
\emph{105}(1-2), 63--80.

\leavevmode\hypertarget{ref-bergelson2016seedlings}{}%
Bergelson, E. (2016a). Bergelson seedlings homebank corpus.
\url{http://doi.org/10.21415/T5PK6D}

\leavevmode\hypertarget{ref-bergelson2016seedlingsdatabrary}{}%
Bergelson, E. (2016b). SEEDLingS corpus. Retrieved January 26, 2018,
from \url{https://nyu.databrary.org/volume/228}

\leavevmode\hypertarget{ref-bergelson2017nature}{}%
Bergelson, E., \& Aslin, R. N. (2017). Nature and origins of the lexicon
in 6-mo-olds. \emph{Proceedings of the National Academy of Sciences},
\emph{114}(49), 12916--12921.

\leavevmode\hypertarget{ref-bergelson2013acquisition}{}%
Bergelson, E., \& Swingley, D. (2013). The acquisition of abstract words
by young infants. \emph{Cognition}, \emph{127}(3), 391--397.

\leavevmode\hypertarget{ref-brent2001role}{}%
Brent, M. R., \& Siskind, J. M. (2001). The role of exposure to isolated
words in early vocabulary development. \emph{Cognition}, \emph{81}(2),
B33--B44.

\leavevmode\hypertarget{ref-colunga2005lexicon}{}%
Colunga, E., \& Smith, L. B. (2005). From the lexicon to expectations
about kinds: A role for associative learning. \emph{Psychological
Review}, \emph{112}(2), 347.

\leavevmode\hypertarget{ref-dale1996lexical}{}%
Dale, P. S., \& Fenson, L. (1996). Lexical development norms for young
children. \emph{Behavior Research Methods, Instruments, \& Computers},
\emph{28}(1), 125--127.

\leavevmode\hypertarget{ref-duncan1989visual}{}%
Duncan, J., \& Humphreys, G. W. (1989). Visual search and stimulus
similarity. \emph{Psychological Review}, \emph{96}(3), 433.

\leavevmode\hypertarget{ref-edelman1998representation}{}%
Edelman, S. (1998). Representation is representation of similarities.
\emph{Behavioral and Brain Sciences}, \emph{21}(4), 449--467.

\leavevmode\hypertarget{ref-firth1957synopsis}{}%
Firth, J. R. (1957). A synopsis of linguistic theory, 1930-1955.
\emph{Studies in Linguistic Analysis}.

\leavevmode\hypertarget{ref-frank2017wordbank}{}%
Frank, M. C., Braginsky, M., Yurovsky, D., \& Marchman, V. A. (2017).
Wordbank: An open repository for developmental vocabulary data.
\emph{Journal of Child Language}, \emph{44}(3), 677--694.

\leavevmode\hypertarget{ref-frank2009development}{}%
Frank, M. C., Vul, E., \& Johnson, S. P. (2009). Development of infants'
attention to faces during the first year. \emph{Cognition},
\emph{110}(2), 160--170.

\leavevmode\hypertarget{ref-fukushima1982neocognitron}{}%
Fukushima, K., \& Miyake, S. (1982). Neocognitron: A self-organizing
neural network model for a mechanism of visual pattern recognition. In
\emph{Competition and cooperation in neural nets} (pp. 267--285).
Springer.

\leavevmode\hypertarget{ref-gelman1986categories}{}%
Gelman, S. A., \& Markman, E. M. (1986). Categories and induction in
young children. \emph{Cognition}, \emph{23}(3), 183--209.

\leavevmode\hypertarget{ref-gleitman2005hard}{}%
Gleitman, L. R., Cassidy, K., Nappa, R., Papafragou, A., \& Trueswell,
J. C. (2005). Hard words. \emph{Language Learning and Development},
\emph{1}(1), 23--64.

\leavevmode\hypertarget{ref-hahn2003similarity}{}%
Hahn, U., Chater, N., \& Richardson, L. B. (2003). Similarity as
transformation. \emph{Cognition}, \emph{87}(1), 1--32.

\leavevmode\hypertarget{ref-harris1954distributional}{}%
Harris, Z. S. (1954). Distributional structure. \emph{Word},
\emph{10}(2-3), 146--162.

\leavevmode\hypertarget{ref-helmholtz1878facts}{}%
Helmholtz, H. (1878). The facts of perception. \emph{Selected Writings
of Hermann Helmholtz}, 1--15.

\leavevmode\hypertarget{ref-james2013principles}{}%
James, W. (1890). \emph{The principles of psychology}. Henry Holt;
Company.

\leavevmode\hypertarget{ref-kemp2005generative}{}%
Kemp, C., Bernstein, A., \& Tenenbaum, J. B. (2005). A generative theory
of similarity. In \emph{Proceedings of the 27th annual conference of the
cognitive science society} (pp. 1132--1137).

\leavevmode\hypertarget{ref-landau1992syntactic}{}%
Landau, B., Smith, L. B., \& Jones, S. (1992). Syntactic context and the
shape bias in children's and adults' lexical learning. \emph{Journal of
Memory and Language}, \emph{31}(6), 807--825.

\leavevmode\hypertarget{ref-macwhinney2000childes}{}%
MacWhinney, B. (2000). \emph{The childes project: The database} (Vol.
2). Psychology Press.

\leavevmode\hypertarget{ref-mandler2000perceptual}{}%
Mandler, J. M. (2000). Perceptual and conceptual processes in infancy.
\emph{Journal of Cognition and Development}, \emph{1}(1), 3--36.

\leavevmode\hypertarget{ref-mareschal2001categorization}{}%
Mareschal, D., \& Quinn, P. C. (2001). Categorization in infancy.
\emph{Trends in Cognitive Sciences}, \emph{5}(10), 443--450.

\leavevmode\hypertarget{ref-pennington2014glove}{}%
Pennington, J., Socher, R., \& Manning, C. (2014). Glove: Global vectors
for word representation. In \emph{Proceedings of the 2014 conference on
empirical methods in natural language processing (emnlp)} (pp.
1532--1543).

\leavevmode\hypertarget{ref-pereira2018toward}{}%
Pereira, F., Lou, B., Pritchett, B., Ritter, S., Gershman, S. J.,
Kanwisher, N., \ldots{} Fedorenko, E. (2018). Toward a universal decoder
of linguistic meaning from brain activation. \emph{Nature
Communications}, \emph{9}(1), 963.

\leavevmode\hypertarget{ref-popov2017target}{}%
Popov, V., \& Reder, L. (2017). Target-to-distractor similarity can help
visual search performance. In \emph{Proceedings of the 39th annual
conference of the cognitive science society} (pp. 968--973).

\leavevmode\hypertarget{ref-quinn1997reexamination}{}%
Quinn, P. C., \& Eimas, P. D. (1997). A reexamination of the
perceptual-to-conceptual shift in mental representations. \emph{Review
of General Psychology}, \emph{1}(3), 271.

\leavevmode\hypertarget{ref-quinn2000emergence}{}%
Quinn, P. C., \& Eimas, P. D. (2000). The emergence of category
representations during infancy: Are separate perceptual and conceptual
processes required? \emph{Journal of Cognition and Development},
\emph{1}(1), 55--61.

\leavevmode\hypertarget{ref-quinn2000understanding}{}%
Quinn, P. C., Johnson, M. H., Mareschal, D., Rakison, D. H., \& Younger,
B. A. (2000). Understanding early categorization: One process or two?
\emph{Infancy}, \emph{1}(1), 111--122.

\leavevmode\hypertarget{ref-rosch1978cognition}{}%
Rosch, E., \& Lloyd, B. B. (1978). Cognition and categorization.

\leavevmode\hypertarget{ref-ILSVRC15}{}%
Russakovsky, O., Deng, J., Su, H., Krause, J., Satheesh, S., Ma, S.,
\ldots{} Fei-Fei, L. (2015). ImageNet Large Scale Visual Recognition
Challenge. \emph{International Journal of Computer Vision (IJCV)},
\emph{115}(3), 211--252. \url{http://doi.org/10.1007/s11263-015-0816-y}

\leavevmode\hypertarget{ref-shepard1970second}{}%
Shepard, R. N., \& Chipman, S. (1970). Second-order isomorphism of
internal representations: Shapes of states. \emph{Cognitive Psychology},
\emph{1}(1), 1--17.

\leavevmode\hypertarget{ref-shwartz2017opening}{}%
Shwartz-Ziv, R., \& Tishby, N. (2017). Opening the black box of deep
neural networks via information. \emph{arXiv Preprint arXiv:1703.00810}.

\leavevmode\hypertarget{ref-sloutsky2003role}{}%
Sloutsky, V. M. (2003). The role of similarity in the development of
categorization. \emph{Trends in Cognitive Sciences}, \emph{7}(6),
246--251.

\leavevmode\hypertarget{ref-sloutsky2001much}{}%
Sloutsky, V. M., Lo, Y.-F., \& Fisher, A. V. (2001). How much does a
shared name make things similar? Linguistic labels, similarity, and the
development of inductive inference. \emph{Child Development},
\emph{72}(6), 1695--1709.

\leavevmode\hypertarget{ref-stager1997infants}{}%
Stager, C. L., \& Werker, J. F. (1997). Infants listen for more phonetic
detail in speech perception than in word-learning tasks. \emph{Nature},
\emph{388}(6640), 381.

\leavevmode\hypertarget{ref-szegedy2016rethinking}{}%
Szegedy, C., Vanhoucke, V., Ioffe, S., Shlens, J., \& Wojna, Z. (2016).
Rethinking the inception architecture for computer vision. In
\emph{Proceedings of the ieee conference on computer vision and pattern
recognition} (pp. 2818--2826).

\leavevmode\hypertarget{ref-trauble2007role}{}%
Träuble, B., \& Pauen, S. (2007). The role of functional information for
infant categorization. \emph{Cognition}, \emph{105}(2), 362--379.

\leavevmode\hypertarget{ref-treisman1988feature}{}%
Treisman, A., \& Gormican, S. (1988). Feature analysis in early vision:
Evidence from search asymmetries. \emph{Psychological Review},
\emph{95}(1), 15.

\leavevmode\hypertarget{ref-tversky1977features}{}%
Tversky, A. (1977). Features of similarity. \emph{Psychological Review},
\emph{84}(4), 327.

\leavevmode\hypertarget{ref-yoshida2005linguistic}{}%
Yoshida, H., \& Smith, L. B. (2005). Linguistic cues enhance the
learning of perceptual cues. \emph{Psychological Science}, \emph{16}(2),
90--95.

\end{document}